\documentstyle[12pt,psfig]{article}
\begin{document}

\begin{center}
{\Large\bf Spherical collapse of a heat conducting fluid
in higher dimensions without horizon}\\[20 mm]

A.~Banerjee and S.~Chatterjee\footnote{Permanent Address :
 New Alipore College, Kolkata  700053, India\\[1mm] Correspondence to: S. Chatterjee,
 email : sujit@juphys.ernet.in}\\
 Relativity and Cosmology Centre,
Jadavpur University, Kolkata - 700032, India
\end{center}

\begin{abstract}
We consider a scenario where the interior spacetime,described by a
heat conducting fluid sphere is matched to a Vaidya metric in
higher dimensions.Interestingly we get a class of solutions, where
following heat radiation the boundary surface collapses without
the appearance of an event horizon at any stage and this happens
with reasonable properties of matter field.The non-occurrence of a
horizon is due to the fact that the rate of mass loss exactly
counterbalanced by the fall of boundary radius.Evidently this
poses a counter example to the so-called cosmic censorship
hypothesis.Two explicit examples of this class of solutions are
also given and it is observed that the rate of collapse is delayed
with the introduction of extra dimensions.The work extends to
higher dimensions our previous investigation in 4D.

\end{abstract}


\bigskip
\section*{1. Introduction}

One of the most intriguing roles in black hole physics and
gravitational collapse theory is played by the occurrence of naked
singularity as a possible end state to several cases of collapsing
objects.So far many cases of gravitational collapse scenarios,
which include models like Vaidya -Papapetro radiation collapse,
Tolman-Bondi inhomogeneous dust spacetime,self similar collapse of
fluid and a few non self-similar situations even,show naked
singularity in course of evolution\cite{Joshi}.These are shell
focusing naked singularity which occur at the centre of the
distribution. In the present case we,however, present an example of a
spherically symmetric heat conducting, inhomogeneous fluid
distribution but with isotropic motion.Such a heat conducting
interior solution has been worked out in a general higher
dimensional spacetime with an arbitrary number of dimensions, as
was originally studied by Santos\cite{santos} in four dimensions.
The matter field has reasonable behaviour satisfying the standard
energy conditions.The interesting feature of such a collapsing
model is that for a particular solution in interior the event
horizon never appears at the boundary surface and the collapse
continues till a singularity is encountered. This singularity may
be said to be naked in the sense that it is not covered by a
horizon.The non-occurrence of the horizon may be explained away by
the fact that the rate of mass loss due to radiation is exactly
balanced by the fall of the boundary radius. In fact we get a
class of this type of solutions and two explicit solutions of this
class are given and discussed.It may be mentioned that we have
studied a similar situation in a recent communication\cite{bc}in
four dimensional spacetime.

The motivation behind this investigation is to primarily see if
dimensionality has any role to play in the appearance of the naked
singularity.The only difference we notice is that the rate of
collapse is delayed when the spacetime is augmented by the
addition of extra spatial dimensions.In section 2 we work out the
mathematical formalism and the paper ends with a discussion.

\section*{2. The Field Equations and Their Integrals}

The general form of the $(n+2)$ dimensional metric is taken as

\begin{eqnarray}
  ds^{2} &=&
  -A^{2}dt^{2}+B^{2}(dr^{2}+r^{2}dX_{n}^{2}).
\end{eqnarray}
where $A=A(r,t)$ and $B=B(r,t)$ and

\begin{equation}
X_{n}^{2}=
d\theta_{1}^{2}+sin^{2}\theta_{1}d\theta_{2}^{2}+......sin^{2}\theta_{1}sin^{2}\theta_{2}sin^{2}\theta_{n-1}d\theta_{n}^{2}
\end{equation}
The energy momentum tensor for a non-viscous heat conducting fluid
reads
\begin{equation}
T_{ij}= (\rho +p)v_{i}v_{j}+pg_{ij}
+q_{i}v_{j}+q_{j}v_{i}
\end{equation}
 where $q^{i}$is the heat flow vector. Taking
into account the fact that the heat flow vector is orthogonal to the
velocity vector we further have, $q_{i}v^{i}=0$.

Using equations (1) and (3) the non-trivial Einstein equations
in co-moving co-ordinates are\cite{bcb}
\begin{eqnarray}
% \nonumber to remove numbering (before each equation)
  G_{00}&=& -\frac{n(n+1)}{2} \frac{\dot{B}^{2}}{B^{2}} +
   \frac{A^{2}}{B^{2}}(n\frac{B''}{B} +
   \frac{n(n-3)}{2}\frac{B'^{2}}{B^{2}}+n^{2}\frac{B'}{Br}) = -\rho A^{2}\\
  G_{11} &=& n\frac{A'B'}{AB} + n\frac{A'}{Ar} +
   \frac{n(n-1)}{2}\frac{B'^{2}}{B^{2}} + n(n-1)\frac{B'}{Br}\nonumber\\
   &+& \frac{B^{2}}{A^{2}}(-n\frac{\ddot{B}}{B}-\frac{n(n-1)}{2}\frac{\dot{B^{2}}}{B^{2}}
   + n\frac{\dot{A}\dot{B}}{AB})= pB^{2}\\
  G_{22} &=& G_{33}= .... ~ G_{nn} = [\frac{A''}{A}+
  (n-1)\frac{B''}{B}+(n-1)\frac{A'}{rA}+\frac{(n-1)(n-4)}{2}\frac{B'^{2}}{B^{2}} \nonumber\\
  &+& (n-1)^{2}\frac{B'}{rB} + (n-2)\frac{A'B'}{AB}-
  \frac{B^{2}}{A^{2}}(n\frac{\ddot{B}}{B}+\frac{n(n-1)}{2}\frac{\dot{B^{2}}}{B^{2}}
  -n\frac{\dot{A}\dot{B}}{AB}=pB^{2} \\
  G_{01} &=& -n\frac{\dot{B'}}{B}+n\frac{\dot{B}B'}{B^{2}}+n\frac{A'\dot{B}}{AB}=-qB^{2}A
\end{eqnarray}
Moreover the isotropy of pressure gives
\begin{equation}
[\frac{A'}{A}+ (n-1)\frac{B'}{B}]'- (\frac{A'}{A}+ \frac{B'}{B})^{2}-\frac{1}{r}[\frac{A'}{A}
+(n-1)\frac{B'}{B}] + 2(\frac{A'}{A})^{2}- (n-2)(\frac{B'}{B})^{2}= 0
\end{equation}
At this stage we like to match the shearfree,spherically symmetric
 line element (1) with an exterior (n+2)-dim. Vaidya metric.In an earlier
 work \cite{cbb} we have obtained a generalised form of the
 Vaidya metric in (n+2)-dim spacetime as
\begin{equation}
ds^{2} = - ( 1-\frac{2m}{\bar{r}}) + 2d\bar{r}dv +{\bar{r}}^{2}(dX_{n})^{2}
\end{equation}
Invoking Israel's junction conditions that at the boundary the first
 and the second fundamental forms calculated from the line elements(1)and (9)
 should be continuous we get \cite{santos} $\bar{r}_{\Sigma}= (rB)_{\Sigma}$
 and $p_{\Sigma}= (qB)_{\Sigma}$.\\ and the mass function calculated
 within the radius $r_{\Sigma}$is given by\cite{bcb}
 \begin{equation}
 m_{\Sigma}= (n-1)[\frac{r^{n+1}(\dot{B})^{2}B^{n-1}}{2A^{2}}- r^{n}B^{n-2}B'-
 \frac{r^{n+1}B'^{2}B^{n-3}}{2}]_{\Sigma}
 \end{equation}
As it is very difficult to solve the field equations in the general
form we choose a special form of the metric coefficient as
\begin{eqnarray}
A=A(r)\\
B=b(r)R(t)
\end{eqnarray}
Now the boundary conditions yield via equation(12) the following
\begin{equation}
2R\ddot{R}+ ( n-1)\dot{R}^{2}+M\dot{R}= 0
\end{equation}
where M and N are constants involving $a_{0},b_{0}$ and their space
derivatives as also 'n' and the suffix zero indicates the
corresponding values on the boundary $r=r_{0}$.\\
It is very difficult to get a general solution of the last equation
in a closed form.A very simple solution
\begin{equation}
R = -ct
\end{equation}
suggests itself and the constant 'c' comes out to be
\begin{equation}
c = \frac{M \pm\sqrt{M^{2}+ 4N(n-1)}}{2(n-1)}
\end{equation}
The choice of the positive signature in equation(15) is needed for the
positive magnitude of the constant c,which in turn selects a collapsing
 system as is evident from the equation(14).Here the range of the time
 coordinate is given by $-\infty< t \leq 0$
 Now the relations(10),(12)and (14) together give the following
 relation
 \begin{equation}
 [\frac{2m}{(n-1)\bar{r}^{n-1}}]_{\Sigma}= 2[\frac{r^{2}b^{2}c^{2}}{2a^{2}}
 -\frac{rb'}{b}-\frac{r^{2}b'^{2}}{2b^{2}}]_{\Sigma}
 \end{equation}
This is a very general result because when $n=2$ we recover
the wellknown four dimensional result obtained by us
 earlier\cite{bc}The equation (16) is quite interesting in
 the sense that at the boundary,for the solution(14) the ratio
 ~~$ \frac{m_{\Sigma}}{\bar{r}^{n-1}}$~~ no longer depends on time.
So one is at liberty to fix up the magnitudes of the arbitrary
constants appearing in (16) to fulfil the above condition.In
this case the equation(16) indicates that the socalled horizon
will never appear at the boundary.So naked singularity is possible
 in this model exhibiting a counter example to the well known
 cosmic censorship hypothesis.\\
 Moreover it has not escaped our notice that the equation (15) gives
 $\frac{dc}{dn}< 0$.The extra dimensions therefore make their
 presence felt in delaying the rate of collapse of the fluid distribution.

\section*{3.~~(n+2) dim. fluid solution with heat flow}
\textbf{Case A :}
If we substitute $x=r^{2}$, the pressure isotropy equation
yields via (13) an important relation
\begin{equation}
\frac{A_{xx}}{A}- (n-1)\frac{F_{xx}}{F}+ 2\frac{A_{x}}{A}\frac{F_{x}}{F}=0
\end{equation}
where $F= 1/B$
If we set $F_{xx}=0$,we get back Maiti's solution\cite{maiti} and for $F_{x}
=0$,Modak's form of solution\cite{modak}in (n+2)dim.as
\begin{equation}
dS^{2}= -(1+\zeta_{0}r^{2})dt^{2}+ R^{2}(t)(dr^{2}+r^{2}X_{n}^{2})
\end{equation}
A striking feature of this line element is that the metric form has
apparently no signature of dimensions.But the number of dimensions
appear explicitly when put in the field equations.Then we get
\begin{eqnarray}
\rho &=& \frac{n(n+1)}{2(1+\zeta_{0}r^{2})^{2}}\frac{1}{t^{2}} \\
p &=& \frac{n}{t^{2}}[\frac{2 \zeta_{0}}{c^{2}(1+\zeta_{0}r^{2})}
-\frac{(n-1)}{2(1+\zeta_{0}r^{2})^{2}}]\\
q &=& -\frac{2n\zeta_{0}r}{(1+\zeta_{0}r^{2})^{2}}\frac{1}{c^{2}t^{3}}
\end{eqnarray}
When $\zeta_{0}=0$, the heat flux vanishes and the line element reduces
to the (n+2)-dim.FRW like spacetime obtained by some of us earlier\cite
{cb}.All the expressions ( 19-21) show that the matter field diverges
at the bigbang singularity of $t=0$ and as we are discussing a
collapsing model$( \dot{R}< 0)$the heat flux is negative,being zero
at the centre of the fluid distribution.It is not difficult to
show that physically realistic situations are achieved when
$\rho> 0, p > 0$ , and  $\rho' < 0$ , $p' < 0$. Skipping mathematical
details one can show that these are satisfied if $c^{2}(n-1)<
2\zeta_{0}$.
Moreover $\rho > p$ ( dominant energy condition) everywhere
further restricts $c^{2}> \frac{2\zeta_{0}}{n}( 1+ \zeta_{0}r^{2})$.
All things put together finally give
\begin{equation}
2\zeta_{0}> c^{2}(n-1) > 2\frac{n-1}{n}\zeta_{0}( 1+\zeta_{0}r^{2})
\end{equation}
which also suggests $\zeta_{0}r_{0}^{2}< \frac{1}{n-1}$
that is
\begin{equation}
\zeta_{0}r_{0}^{2} < 1
\end{equation}
For a heat conducting fluid another stringent requirement is
\cite{Kolassis} $| \rho + p~ | > 2 | q |$,
where $q = (g_{ij}q^{i}q^{j})^{\frac{1}{2}}$. Skipping
further mathematical details it is not difficult to show that the
 above inequality leads to
\begin{equation}
( 1-\frac{2\zeta_{0}r}{c})^{2} > -2\frac{\zeta_{0}}{c^{2}}(1 - \zeta_{0}r^{2})
\end{equation}
which is always satisfied in view of equation (23).\\
\textbf{Case B :}
In this section we search for a new type of solution for the key equation (17)
where the metric dependence on the extra dimensions are more explicit.
Here we assume $ A_{xx} = 0 $
Avoiding intermediate mathematical steps for economy of space one gets a
new line element as
\begin{equation}
ds^{2}= - ( 1+\zeta_{0}r^{2})^{2}dt^{2}
+ \frac{R^{2}(t)}{(1 + \zeta_{0}r^{2})^{\frac{n+1}{n-1}}+a }(dr^{2}+r^{2}dX_{n}^{2})
\end{equation}
Here also, if we set $\zeta_{0} = 0 $ the heat flow vanishes and we get back
the (n+2)- dimensional FRW type of metric. Evidently the presence of the extra
dimensions is more explicit in equation (25) compared to equation (18).

Since the solutions (25) are examples of the classes belonging to the
expressions (11) and (12) they along with explicit form of R(t) as assumed in (14)
 again represent collapse with horizon not appearing at the boundary at
 any stage of collapse, which means that the end state is again a naked
 singularity.

 Setting a=0, for simplicity we get
 \begin{eqnarray}
 % \nonumber to remove numbering (before each equation)
   \rho &=& \frac{n}{( 1+\zeta_{0}r^{2})t^{2}}[\frac{n+1}{2}
   + \frac{2(n+1)}{n-1}\frac{\zeta_{0}}{c^{2}}\{(n+1)-
   2\zeta_{0}r^{2}\}\{(1+\zeta_{0}r^{2})^{\frac{2(n+1)}{n-1}} \\
   p &=& \frac{n(1+\zeta_{0}r^{2})}{c^{2}t^{2}}[2\zeta_{0}(\zeta_{0}r^{2}-n)\
    -\frac{n-1}{2}c^{2}(1+ \zeta_{0}r^{2})]^{\frac{-2(n+1)}{n-1}} \\
   q &=& -\frac{2n\zeta_{0}r}{c^{2}t^{3}}( 1+ \zeta_{0}r^{2})^{\frac{n-1}{4}}
 \end{eqnarray}
So as $t\rightarrow 0$ the density, pressure and heat flux all diverge.
\section{Discussion}

In this work we have studied a very idealised model of spherically
symmetric object radiating away its mass with constant luminosity
where the material particles undergo non-geodetic but shear-free
motion while transporting heat to the surface.While we have not
been able to find out an equation of state,the matter field
satisfies all the three energy conditions.The most interesting
finding is the absence of any event horizon in our class of
models so that naked singularity may eventually occur.This can
be explained away by the fact that here both the mass function
and the area radius of the radiating sphere vary linearly with
time so that the ratio is static.Put it otherwise the collapsing
sphere does not admit any accumulation of energy at any stage
because as much energy is being radiated as it generates.

As a future exercise one should investigate the whole
situation in a causal model in the framework of socalled
extended thermodynamics.


\begin{thebibliography}{22}
\bibitem{Joshi} P.S.~Joshi,{\it Global Aspects in Gravitation and Cosmology}
(Clarendon,Oxford, 1993).
\bibitem{santos} N.O.~Santos,{\it Mon. Not.R. Astron. Soc.} \textbf{216}, 403 (1985).
\bibitem{bc} A.~Banerjee, S.~Chatterjee and N.~Dadhich,{\it Mod.Phys.Lett.A} \textbf{35},2335~(2002).
\bibitem{bcb} B.~Bhui,S.~Chatterjee and A.~Banerjee,{\it Astrophys and Space Sci.} \textbf{226}
 ,7 (1995)
\bibitem{cbb} S.~Chatterjee, B.~Bhui and A.~Banerjee, {\it J.Math.Phys.}\textbf{31}, 2208 (1990).
\bibitem{maiti}S.R.~Maiti,{\it Phys.Rev.} D \textbf{25},2518 ( 1982 ).
\bibitem{modak} B.~Modak,{\it J.Astrophys. Astron.} \textbf{5}, 317 (1984).
\bibitem{cb} S.~Chatterjee and B.~Bhui, {\it Mon. Not. R. Astron. Soc.}\textbf{247}, 577 (1990).
\bibitem{Kolassis} C.~Kolassis,N.~Santos and D.~Tsoubelis, {\it Class.Quantum Grav.}
\textbf{5} ( 1329).





\end{thebibliography}
\end{document}